\documentclass[12pt]{article}
\usepackage{graphicx}
\begin{document}
\pagenumbering{arabic}
\begin{titlepage}
\title{Gravitational waves and the breaking of parallelograms
in space-time}

\author{J. W. Maluf$\,^{(1,a)}$, S. C. Ulhoa$\,^{(2,b)}$ 
and J. F. da Rocha-Neto$\,^{(1,c)}$}

\date{}
\maketitle

{\footnotesize
\noindent{  (1) Instituto de F\'{\i}sica, Universidade de 
Bras\'{\i}lia, C.P. 04385, 70.919-970 Bras\'{\i}lia DF, Brazil}\par
\bigskip
\noindent{ (2) Faculdade UnB Gama, Universidade de Bras\'ilia,
72.405-610, Gama DF, Brazil}}
\bigskip

\begin{abstract}
We show that plane-fronted gravitational waves induce the breaking of
parallelograms in space-time, in the context of the teleparallel equivalent
of general relativity (TEGR). The breaking of parallelograms can be 
shown by considering a thought experiment that consists of a simple 
physical configuration, similar to the experimental setup that is expected to
lead to the measurement of gravitational waves with the use of laser 
interferometers. An incident beam of light splits into two beams running 
along perpendicular arms, endowed with fixed mirrors at the extremes. The 
reflected light beams are detected at the same point of the splitting. Along 
each arm, the two light beams define two null 
vectors: the forward vector and the reflected vector. We show that the sum of
these four vectors, the forward and reflected null vectors along the two arms,
do form a parallelogram in flat space-time, but not in the presence of 
plane-fronted gravitational waves. The non-closure of the parallelogram 
is a manifestation of the torsion of the space-time, and in this context
indicates the existence of gravitational waves.
\end{abstract}
\thispagestyle{empty}
\vfill
\noindent PACS numbers: 04.20.-q, 04.20.Cv, 04.30.-w\par

\bigskip
{\footnotesize
\noindent (a) wadih@unb.br, jwmaluf@gmail.com\par
\noindent (b) sc.ulhoa@gmail.com\par
\noindent (c) rocha@fis.unb.br\par}

\end{titlepage}
\bigskip
\section{Introduction}
The dynamics of the gravitational field may be described either in terms of
the metric and curvature tensors, or in terms of the tetrad fields and the 
torsion tensor. The latter approach is known as teleparallel gravity. The 
reformulation of Einstein's theory in terms of tetrad fields and of the 
torsion tensor is the teleparallel equivalent of general relativity (TEGR) 
\cite{Hay,Hehl2,Maluf4,Maluf7,FG}. 
The field equations for the gravitational field in the standard metric 
formulation and in the TEGR are essentially the same. However, 
different geometrical frameworks naturally lead to different approaches and 
forms of investigation of the theory. In particular, in the TEGR there are
definitions and physical results that are expressed in terms of the torsion
tensor.

Gravitational waves are one of the most important consequences of general 
relativity. They are due to the dynamical nature of space-time, which we 
consider to be described by Einstein's equations. It is natural to consider a
gravitational wave as a kind of ripple in the background flat geometry. 
Gravitational waves may be classified either as non-linear waves, which are 
exact solutions of Einstein's equations, or linearized waves,
which are solutions of the linearized Einstein's equations, presented in 
standard textbooks on general relativity. One of the simplest realizations of
a non-linear gravitational wave is given by the solution known as plane-fronted
gravitational wave, studied by Ehlers and Kundt \cite{Ehlers}.

The present attempts to observe gravitational waves are based on laser
interferometers, constituted by two long and perpendicular arms that are able,
in principle, to detect gravitational waves travelling in the direction normal
the plane formed by the two arms (see, for instance, Refs. 
\cite{Living1,Living2}).  The mirrors at the end of the long arms may be 
attached to test masses that are hung from wires, and are free to swing in the 
horizontal directions, or may be fixed at the end of the arms. In the present
setup, the mirrors are fixed at the end of the arms. 

In this article, we will consider a thought experiment that consists in laser
beams travelling back and forth along
perpendicular arms, in the presence of a plane-fronted gravitational wave 
travelling in the $z$ direction. Let us denote by $v^\mu$ and $w^\mu$ the null 
vectors that represent, respectively, (i) the laser beam trajectory from the 
splitting point $S$ to the fixed mirror $M_1$, and (ii) from $M_1$ back to $S$,
along the $x$ direction. Along the $y$ 
direction, the vectors $a^\mu$ and $b^\mu$ travel (iii) from the splitting point 
$S$ to the fixed mirror $M_2$, and (iv) from $M_2$ back to $S$, respectively. 
In flat space-time, these four null vectors form a parallelogram in the sense 
that $v^\mu +w^\mu = a^\mu + b^\mu$. We will show in this analysis that
in the presence of a plane-fronted gravitational wave, these null vectors
no longer form a parallelogram, because $v^\mu +w^\mu \ne a^\mu + b^\mu$. This
fact is an indication that the space-time has torsion. By establishing the
tetrad frame adapted to stationary observers in the space-time of plane-fronted
gravitational waves, we will obtain the torsion tensor that 
precisely explains the breaking of the parallelogram. The result described here
is geometrically similar to the breaking of parallelograms in the Schwarzschild
space-time, in the context of the Pound-Rebka experiment 
\cite{Schucking,Maluf1}. The breaking of parallelograms shows that torsion 
is an intrinsic geometric property of space-time, and indicates the presence 
of gravitational waves. In the present approach, the existence of gravitational
waves is verified by the lapse of time between the arrival of the two reflected
(perpendicular) light beams, as we will see.

The article is organized as follows. In Section 2 we recall the interpretation
of tetrad fields as frames adapted to arbitrary observers in space-time. In 
Section 3 the TEGR is briefly presented. The breaking of parallelograms in
the space-time of plane-fronted gravitational waves is discussed in Sections 4
and 5. In Section 6 we present our conclusions.

Notation: space-time indices $\mu, \nu, ...$ and SO(3,1) indices
$a, b, ...$ run from 0 to 3. Time and space indices are indicated
according to $\mu=0,i,\;\;a=(0),(i)$. The tetrad fields are denoted
$e^a\,_\mu$, and the torsion tensor reads
$T_{a\mu\nu}=\partial_\mu e_{a\nu}-\partial_\nu e_{a\mu}$.
The flat, Minkowski space-time metric tensor raises and lowers
tetrad indices and is fixed by
$\eta_{ab}=e_{a\mu} e_{b\nu}g^{\mu\nu}=diag(-1,1,1,1)$. The 
determinant of the tetrad fields is represented by 
$e=\det(e^a\,_\mu)$.

The torsion tensor defined above is often related to the object of
anholonomity $\Omega^\lambda\,_{\mu\nu}$ via 
$\Omega^\lambda\,_{\mu\nu}= e_a\,^\lambda T^a\,_{\mu\nu}$.
However, we assume that the space-time geometry is defined by the 
tetrad fields only, and in this case the only possible non-trivial definition
for the torsion tensor is given by $T^a\,_{\mu\nu}$. This torsion tensor is 
related to the antisymmetric part of the Weitzenb\"ock  connection 
$\Gamma^\lambda_{\mu\nu}=e^{a\lambda}\partial_\mu e_{a\nu}$, which
establishes the Weitzenb\"ock space-time. The curvature of the Weitzenb\"ock
connection vanishes. However, the tetrad fields also yield the metric tensor,
which establishes the Riemannian geometry. Therefore in the framework of 
a geometrical theory based only on tetrad fields, one may use the 
concepts of both Riemannian and Weitzenb\"ock geometries. 

\section{Tetrad fields as reference frames}

We recall the discussion presented in Refs. \cite{Maluf1-1,Maluf2-2} regarding
the characterization of tetrad fields as reference frames in space-time. A 
frame may be characterized in a coordinate invariant way by the inertial 
accelerations, represented by the acceleration tensor. 

We denote by $x^\mu(s)$ the worldline $C$ of an observer in 
space-time, where $s$ is the proper time of the observer. The velocity
of the observer on $C$ reads $u^\mu=dx^\mu/ds$. The observer's velocity 
is identified with the $a=(0)$ component of $e_a\,^\mu$:
$u^\mu(s)=e_{(0)}\,^\mu$ (we are assuming $c=1$ for the speed of light).
The acceleration $a^\mu$ of the observer
is given by the absolute derivative of $u^\mu$ along $C$ \cite{Hehl},

\begin{equation}
a^\mu= {{Du^\mu}\over{ds}} ={{De_{(0)}\,^\mu}\over {ds}} =
u^\alpha \nabla_\alpha e_{(0)}\,^\mu\,, 
\label{1}
\end{equation}
where the covariant derivative is constructed out of the Christoffel 
symbols. Thus, $e_a\,^\mu$ and its derivatives determine the velocity and
acceleration of an observer along the worldline. The set of tetrad fields for
which $e_{(0)}\,^\mu$ describe a congruence of timelike curves is adapted to 
a class of observers characterized by the velocity field 
$u^\mu=e_{(0)}\,^\mu$ and by the acceleration $a^\mu$. 

We may consider not only the acceleration of observers along
trajectories whose tangent vectors are given by $e_{(0)}\,^\mu$, but
the acceleration of the whole frame along $C$. The acceleration of the frame
is determined by the absolute derivative of $e_a\,^\mu$ along the path 
$x^\mu(s)$. Thus, assuming that the observer carries an orthonormal tetrad 
frame $e_a\,^\mu$, the acceleration of the latter along the path is given by 
\cite{Mashh}

\begin{equation}
{{D e_a\,^\mu} \over {ds}}=\phi_a\,^b\,e_b\,^\mu\,,
\label{2}
\end{equation}
where $\phi_{ab}$ is the antisymmetric acceleration tensor. According to Ref.
\cite{Mashh}, in analogy with the Faraday tensor we may identify 
$\phi_{ab} \rightarrow ({\bf a}, {\bf \Omega})$, where 
${\bf a}$ is the translational acceleration ($\phi_{(0)(i)}=a_{(i)}$)
and ${\bf \Omega}$ is the angular velocity of the local spatial frame with 
respect to a non-rotating (Fermi-Walker transported) frame. It follows that

\begin{equation}
\phi_a\,^b= e^b\,_\mu {{D e_a\,^\mu} \over {ds}}=
e^b\,_\mu \,u^\lambda\nabla_\lambda e_a\,^\mu\,.
\label{3}
\end{equation}

Therefore, given any set of tetrad fields for an arbitrary gravitational field
configuration, its geometrical interpretation may be obtained by interpreting
$e_{(0)}\,^\mu=u^\mu$, where $u^\mu$ is the velocity of the observer, and by 
the values of the acceleration tensor $\phi_{ab}$. 

The acceleration vector 
$a^\mu$ defined by Eq. (1) may be projected on a frame in order to yield

\begin{equation}
a^b= e^b\,_\mu a^\mu=e^b\,_\mu u^\alpha \nabla_\alpha
e_{(0)}\,^\mu=\phi_{(0)}\,^b\,.
\label{4}
\end{equation}
Thus, $a^\mu$ and $\phi_{(0)(i)}$ are not different accelerations of the 
frame. The acceleration $a^\mu$ may be rewritten as

\begin{eqnarray}
a^\mu&=& u^\alpha \nabla_\alpha e_{(0)}\,^\mu 
=u^\alpha \nabla_\alpha u^\mu =
{{dx^\alpha}\over {ds}}\biggl(
{{\partial u^\mu}\over{\partial x^\alpha}}
+\,^0\Gamma^\mu_{\alpha\beta}u^\beta \biggr) \nonumber \\
&=&{{d^2 x^\mu}\over {ds^2}}+\,^0\Gamma^\mu_{\alpha\beta}
{{dx^\alpha}\over{ds}} {{dx^\beta}\over{ds}}\,,
\label{5}
\end{eqnarray}
where $^0\Gamma^\mu_{\alpha\beta}$ are the Christoffel symbols.
If $u^\mu=e_{(0)}\,^\mu$ represents a geodesic
trajectory, then the frame is in free fall and 
$a^\mu=0=\phi_{(0)(i)}$. Therefore we conclude that nonvanishing
values of $\phi_{(0)(i)}$ represent inertial accelerations
of the frame.

Following Ref. \cite{Maluf1-1}, we take into account the 
orthogonality of the tetrads and write Eq. (3) as
$\phi_a\,^b= -u^\lambda e_a\,^\mu \nabla_\lambda e^b\,_\mu$, 
where $\nabla_\lambda e^b\,_\mu=\partial_\lambda e^b\,_\mu-
\,^0\Gamma^\sigma_{\lambda \mu} e^b\,_\sigma$. Next we consider
the identity $\partial_\lambda e^b\,_\mu-
\,^0\Gamma^\sigma_{\lambda \mu} e^b\,_\sigma+\,\,
^0\omega_\lambda\,^b\,_c e^c\,_\mu=0$, where
$^0\omega_\lambda\,^b\,_c$ is the metric compatible 
Levi-Civita connection given by

$$^o\omega_{\mu ab}=-{1\over 2}e^c\,_\mu(
\Omega_{abc}-\Omega_{bac}-\Omega_{cab})\;,$$

$$\Omega_{abc}=e_{a\nu}(e_b\,^\mu\partial_\mu e_c\,^\nu-
e_c\,^\mu\partial_\mu e_b\,^\nu)\;,$$
and express $\phi_a\,^b$ according to

\begin{equation}
\phi_a\,^b=e_{(0)}\,^\mu(\,\,^0\omega_\mu\,^b\,_a)\,.
\label{6}
\end{equation}
Finally we take into account the identity 
$\,\,^0\omega_\mu\,^a\,_b= -K_\mu\,^a\,_b$, where $K_\mu\,^a\,_b$ 
is the contorsion tensor defined by

\begin{equation}
K_{\mu ab}={1\over 2}e_a\,^\lambda e_b\,^\nu(T_{\lambda \mu\nu}+
T_{\nu\lambda\mu}+T_{\mu\lambda\nu})\,,
\label{7}
\end{equation}
and $T_{\lambda \mu\nu}=e^a\,_\lambda T_{a\mu\nu}$. 
After simple manipulations we arrive at

\begin{equation}
\phi_{ab}={1\over 2} \lbrack T_{(0)ab}+T_{a(0)b}-T_{b(0)a}
\rbrack\,.
\label{8}
\end{equation}

The expression above is not invariant under local SO(3,1) 
transformations, and for this reason the values of $\phi_{ab}$
may characterize the frame. However, Eq. (8) is invariant under coordinate
transformations. We interpret $\phi_{ab}$ as the inertial accelerations of the
frame. An alternative (kinematic) characterization of the frame consists in 
(i) identifying, as above, the timelike component $e_{(0)}\,^\mu$ of the tetrad
field with the velocity $u^\mu$ of the observer, i.e., 
$e_{(0)}\,^\mu=u^\mu$, and (ii) specifying the orientation of the spacelike
components $e_{(i)}\,^\mu$ (along the standard unit vectors of orientation at
spatial infinity, for instance). In both forms of characterization, six 
components of the tetrad field are fixed.

In Ref. \cite{Maluf1-1} we applied definition (8) to the analysis
of two simple configurations of tetrad fields in the flat Minkowski
spacetime. We considered the frame adapted to linearly accelerated
observers, and to a stationary frame whose four-velocity is
$e_{(0)}\,^\mu=(1,0,0,0)$ and which rotates around the $z$ axis.
The components $\phi_{(0)(i)}$ and $\phi_{(i)(j)}$ yield the 
known and expected values of the translational acceleration and of the 
angular velocity of the frame, respectively. 

\section{The teleparallel equivalent of general relativity}

We are assuming that the space-time geometry is defined by the tetrad fields
only. In this case the only possible definition for the torsion
tensor is given by $T_{a\mu\nu}=\partial_\mu e_{a\nu}-\partial_\nu e_{a\mu}$.
The equivalence of the TEGR with Einstein's general relativity may be
understood by means of an identity between the scalar curvature 
$R(e)$ constructed out of the tetrad fields, and a combination of
quadratic terms of the torsion tensor, 

\begin{equation}
eR(e)\equiv -e({1\over 4}T^{abc}T_{abc}+{1\over 2}T^{abc}
T_{bac}-T^aT_a)
+2\partial_\mu(eT^\mu)\,.
\label{9}
\end{equation}
The formulation of Einstein's general relativity in the context of the
teleparallel geometry is discussed in Refs.
\cite{Hehl2,Maluf4,Maluf7,FG,Hehl1}. The Lagrangian density of the TEGR is 
given by the combination of the quadratic terms on the right hand side of 
eq. (9),

\begin{eqnarray}
L&=& -k e({1\over 4}T^{abc}T_{abc}+{1\over 2}T^{abc}T_{bac}-
T^aT_a) -L_M\nonumber \\
&\equiv& -ke\Sigma^{abc}T_{abc}-L_M\,, 
\label{10}
\end{eqnarray}
where $k=c^3/16\pi G$, $T_a=T^b\,_{ba}$, 
$T_{abc}=e_b\,^\mu e_c\,^\nu T_{a\mu\nu}$ and

\begin{equation}
\Sigma^{abc}={1\over 4} (T^{abc}+T^{bac}-T^{cab})
+{1\over 2}( \eta^{ac}T^b-\eta^{ab}T^c)\;.
\label{11}
\end{equation}
$L_M$ stands for the Lagrangian density of the matter fields.
The field equations derived from (10) are equivalent to 
Einstein's equations. They read

\begin{equation}
e_{a\lambda}e_{b\mu}\partial_\nu (e\Sigma^{b\lambda \nu} )-
e (\Sigma^{b\nu}\,_aT_{b\nu\mu}-
{1\over 4}e_{a\mu}T_{bcd}\Sigma^{bcd} )={1\over {4k}}eT_{a\mu}\,,
\label{12}
\end{equation}
where
$\delta L_M / \delta e^{a\mu}=eT_{a\mu}$. 
It is possible to show that the left hand side of the equation
above may be rewritten as
${1\over 2}e\left[ R_{a\mu}(e)-{1\over 2}e_{a\mu}R(e)\right]$, which proves
the equivalence of the present formulation with Einstein's theory.

Equation (12) may be simplified as 

\begin{equation}
\partial_\nu(e\Sigma^{a\lambda\nu})={1\over {4k}}
e\, e^a\,_\mu( t^{\lambda \mu} + T^{\lambda \mu})\;,
\label{13}
\end{equation}
where $T^{\lambda\mu}=e_a\,^{\lambda}T^{a\mu}$ and
$t^{\lambda\mu}$ is defined by

\begin{equation}
t^{\lambda \mu}=k(4\Sigma^{bc\lambda}T_{bc}\,^\mu-
g^{\lambda \mu}\Sigma^{bcd}T_{bcd})\,.
\label{14}
\end{equation}
In view of the antisymmetry property 
$\Sigma^{a\mu\nu}=-\Sigma^{a\nu\mu}$ it follows that

\begin{equation}
\partial_\lambda
\left[e\, e^a\,_\mu( t^{\lambda \mu} + T^{\lambda \mu})\right]=0\,.
\label{15}
\end{equation}
The equation above yields the continuity (or balance) equation,

\begin{equation}
{d\over {dt}} \int_V d^3x\,e\,e^a\,_\mu (t^{0\mu} +T^{0\mu})
=-\oint_S dS_j\,
\left[e\,e^a\,_\mu (t^{j\mu} +T^{j\mu})\right]\,.
\label{16}
\end{equation}
We identify
$t^{\lambda\mu}$ as the gravitational energy-momentum tensor
\cite{Maluf7}, and

\begin{equation}
P^a=\int_V d^3x\,e\,e^a\,_\mu (t^{0\mu} 
+T^{0\mu})\,,
\label{17}
\end{equation}
as the total energy-momentum contained within a volume $V$ of the
three-dimensional space. In view of (13), Eq. (17) may be written as 

\begin{equation}
P^a=-\int_V d^3x \partial_j \Pi^{aj}\,,
\label{18}
\end{equation}
where $\Pi^{aj}=-4ke\,\Sigma^{a0j}$, which is the momentum
canonically conjugated to $e_{aj}$ \cite{Maluf4,Maluf7}.
The expression above is the 
definition for the gravitational energy-momentum discussed in Refs.
\cite{Maluf3,Maluf7}, obtained in the framework of the Hamiltonian vacuum
field equations. We note that (15) is a true energy-momentum 
conservation equation.

The emergence of a nontrivial total divergence is a feature of 
theories with torsion. The integration of this total divergence yields a
surface integral. If we consider the $a=(0)$ component of Eq. (18) and 
adopt asymptotic boundary conditions for the tetrad fields, we find 
\cite{Maluf3} that the resulting expression is precisely the surface 
integral at infinity that defines the ADM energy \cite{ADM}. This fact 
is a strong indication that Eq. (10) does indeed represent the gravitational
energy-momentum.


\section{The breaking of parallelograms in space-time}

We will consider the space-time of a plane-fronted gravitational wave.  The
space-time is described by a line element that is an exact solution of 
Einstein's equations, and is given by \cite{Ehlers,HS}

\begin{equation}
ds^2=dx^2+dy^2+2du\,dv+H(x,y,u)du^2\,.
\label{19}
\end{equation}
As in Section 2, we assume $c=1$. The function $H(x,y,u)$ satisfies

\begin{equation}
\biggl( {\partial^2 \over {\partial x^2}}+
{\partial^2 \over {\partial y^2}}\biggr)H(x,y,u)=0\,.
\label{20}
\end{equation}
Transforming the $(u,v)$ to $(t,z)$ coordinates, where

$$u={1\over \sqrt{2}}(z-t)\,, \ \ \ \  v={1\over \sqrt{2}}(z+t)\,,$$
we find

\begin{equation}
ds^2=\biggl({H\over 2} -1\biggr)dt^2+dx^2+dy^2+
\biggl({H\over 2}+1\biggr) dz^2-H\,dt dz\,.
\label{21}
\end{equation}
The function $H$ is required to satisfy only Eq. (20). However,
it would be interesting to specify $H$ such that it describes 
a wave-packet  \cite{HS}. 

Before we investigate the action of a plane-fronted gravitational wave in
the experimental setup, let us establish the physical configuration first 
in flat space-time, according to the discussion in Section 1.
We will consider a laser beam in the $xy$ plane that splits
into two beams: one beam along an arm of proper length $L$, in the $x$ 
direction, and another beam along a similar arm, also with length $L$, in 
the $y$ direction. The splitting point is denoted by $S$, and has coordinates
$(x_0,y_0)$. At the ends of each arm there are fixed mirrors, $M_1$ and 
$M_2$, that reflect the light beams back to the splitting point $S$.
The experimental setup is displayed in Fig. 1. It may be characterized as
a parallelogram in the $(+x,+y)$ directions. In Section 5 we will also consider
the $(+x,-x)$ and $(+x,-y)$ directions.

\begin{figure}[h]
\centering
\includegraphics[width=1.00\textwidth]{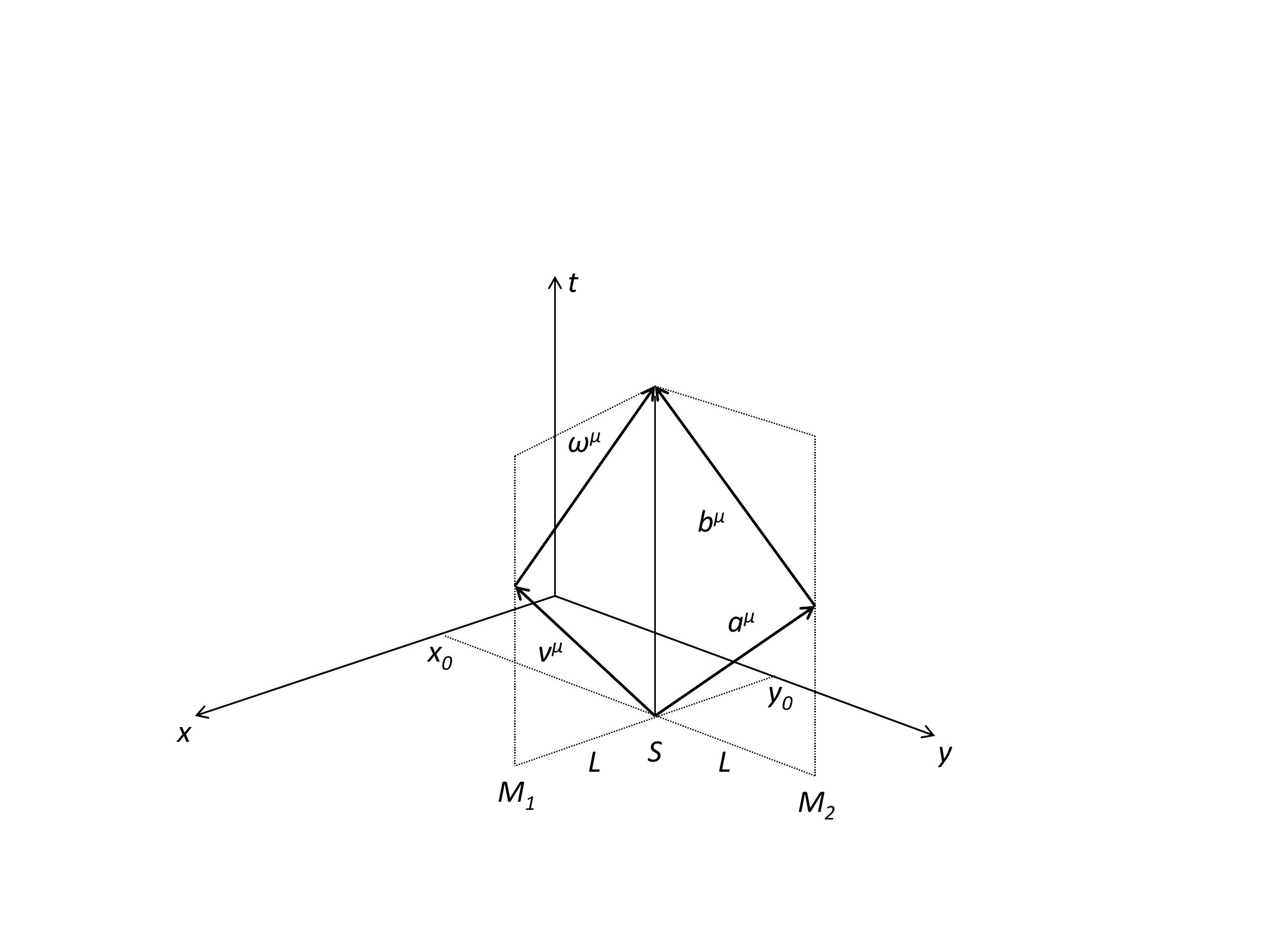}
\caption{The parallelogram in the $(+x,+y)$ directions}
\end{figure}

The trajectory of a light ray in the $x$ direction is described by a null 
vector $v^\mu(t,x,y,z)=(v^0,v^1,0,0)$ that satisfies the condition 
$v^\mu v^\nu g_{\mu\nu}=0$. For a light ray travelling in the positive 
direction along the $x$ axis, the latter condition yields 
$v^1=(-g_{00}/g_{11})^{1/2}\, v_0=(1-H/2)^{1/2}\,v_0$. We will fix the value of
$v^1$ by means of the following procedure. For a vector $v^\mu$ that represents
a light ray between the point $S$ and the position $M_1$ of the mirror, we 
require $v^1$ to have length $L$, in the presence of a plane fronted 
gravitational wave. Thus we require

\begin{equation}
v^1=\int_{x_0}^{x_0+L}dx\,\sqrt{g_{11}}=L\,.
\label{22}
\end{equation}
It follows that 

\begin{equation}
v^0={L\over{(1-H/2)^{1/2}}}\,, \ \ \ \ \ H=H(x^0+L,y_0,t_0+dt)\,,
\label{23}
\end{equation}
where $t_0$ is the time coordinate of the emitted light ray at the point $S$, 
and $t_0+dt$ is the time coordinate when the light ray reaches $M_1$. In flat
space-time, the reflected light ray returns to the point $S$ in the time
coordinate $t_0+2dt$. The values $t_0$, $t_0+dt$ and $t_0+2dt$ parametrize the
physical events also in the presence of a plane-fronted gravitational wave.
Therefore the vector $v^\mu$ is defined by 

\begin{equation}
v^\mu= \biggl( {L\over{(1-H/2)^{1/2}}},L,0,0 \biggl)\,, \ \ \ \ \
H=H(x_0+L,y_0,t_0+dt)\,.
\label{24}
\end{equation}
We define the reflected vector $w^\mu$, that travels from $M_1$ back to $S$, 
according to 

\begin{equation}
w^\mu= \biggl( {L\over{(1-H/2)^{1/2}}},-L,0,0 \biggl)\,, \ \ \ \ \
H=H(x_0,y_0,t_0+2dt)\,.
\label{25}
\end{equation}
Expression (22) for the component $v_1$ holds only for a plane-fronted 
gravitational wave. Note that in the context of a linearized wave, $v_1$ would
not be given by Eq. (22); it would be time dependent.

Along the $y$ direction, the vectors $a^\mu$ and $b^\mu$ that travel from $S$ 
to the mirror $M_2$, and from $M_2$ back to $S$, are constructed in similarity 
to $v^\mu$ and $w^\mu$. They read

\begin{equation}
a^\mu= \biggl( {L\over{(1-H/2)^{1/2}}},0,L,0 \biggl)\,, \ \ \ \ \
H=H(x_0,y_0+L,t_0+dt)\,,
\label{26}
\end{equation}

\begin{equation}
b^\mu= \biggl( {L\over{(1-H/2)^{1/2}}},0,-L,0 \biggl)\,, \ \ \ \ \
H=H(x_0,y_0,t_0+2dt)\,.
\label{27}
\end{equation}

The vectors $v^\mu$, $w^\mu$, $a^\mu$ and $b^\mu$ are defined for a fixed 
coordinate position $z_0$. They are located at $(x_0+L,y_0,t_0+dt)$,
$(x_0,y_0,t_0+2dt)$, $(x_0,y_0+L,t_0+dt)$ and $(x_0,y_0,t_0+2dt)$,
respectively. We find it suitable to depict them as in Fig. 1,
since they do form a parallelogram in a flat geometry.

Let us define the breaking of the parallelogram in space-time by the vector

\begin{equation}
\Delta^\mu=(v^\mu+w^\mu)-(a^\mu+b^\mu)\,.
\label{28}
\end{equation}
It is easy to see that for $i=1,2,3$ we have $\Delta^i=0$, and that for 
$\Delta^0$ we have

\begin{eqnarray}
\Delta^0&=& L \biggl\{ 
{1\over {[1-{1\over 2}H(x_0+L,y_0,t_0+dt)]^{1/2}}}  \nonumber \\
&{}&+{1\over {[1-{1\over 2}H(x_0,y_0,t_0+2dt)]^{1/2}}} \nonumber \\
&{}&-{1\over {[1-{1\over 2}H(x_0,y_0+L,t_0+dt)]^{1/2}}} \nonumber \\
&{}&-{1\over {[1-{1\over 2}H(x_0,y_0,t_0+2dt)]^{1/2}}}
\biggl\}\,.
\label{29}
\end{eqnarray}
The second and fourth terms above cancel each other, and we are left with

\begin{equation}
\Delta^0=L  
{\{[1-{1\over 2}H(x_0,y_0+L,t_0+dt)]^{1/2}-
[1-{1\over 2}H(x_0+L,y_0,t_0+dt)]^{1/2}\}
\over
[1-{1\over 2}H(x_0+L,y_0,t_0+dt)]^{1/2}
[1-{1\over 2}H(x_0,y_0+L,t_0+dt)]^{1/2}}\,.
\label{30}
\end{equation}

The only assumption we make in this analysis is that $L$ is sufficiently small,
so that we can expand $H$ in terms of $L$ according to 

\begin{equation}
H(x_0+L,y_0,t_0)\approx H(x_0,y_0,t_0)+
{{\partial H(x_0,y_0,t_0)}\over {\partial x}}L\equiv
H+{{\partial H}\over{\partial x}} L\,.
\label{31}
\end{equation}
Likewise,

\begin{equation}
H(x_0,y_0+L,t_0)\approx H(x_0,y_0,t_0)+
{{\partial H(x_0,y_0,t_0)}\over {\partial y}}L\equiv
H+{{\partial H}\over{\partial y}} L\,.
\label{32}
\end{equation}
We may consider the function $H$ to represent a wave packet in the form of a
Gaussian function, that travels in the $z$ direction. The approximation above 
holds if $L$ is much smaller compared to the width of the Gaussian (or if $H$
varies very weakly over the space region of Fig. 1). In view of the 
approximation above, Eq. (30) is finally simplified to

\begin{equation}
\Delta^0={L^2\over 4}  \biggl[
{{ ( {{\partial H}\over {\partial x}})
-( {{\partial H}\over {\partial y}})}
\over
{  (1-{1\over 2}H)^{3/2}  }} 
\biggr]\,.
\label{33}
\end{equation}
The equation above expresses the non-closure in space-time of the 
parallelogram displayed in Fig. 1, and thus indicates the presence of torsion 
in the space-time geometry.

We will present here the torsion tensor that precisely explains the 
breaking of the parallelogram of Fig. 1. For this purpose, we need to 
establish the reference frame for the measurement of $\Delta^0$. 
This reference frame is the same one adopted in Ref. \cite{Maluf2}, in the
analysis of the energy-momentum of plane-fronted gravitational waves. As in
Section 2, we denote by $x^\mu(s)$ the worldline $C$ of an arbitrary observer
in space-time, and by $u^\mu=dx^\mu/ds$ its velocity along $C$. 
We identify the timelike component $e_{(0)}\,^\mu$ of the inverse tetrad field 
with the velocity of the observer, i.e., $e_{(0)}\,^\mu=u^\mu$.
The components $e_{(i)}\,^\mu$,
for $i=1,2,3$ may be specified by requiring these spacelike vectors to be 
oriented asymptotically along the three Cartesian axes 
$x,y,z$. The tetrad frame established in Ref. \cite{Maluf2} is 
well suited for our purposes.

We will consider the stationary observer determined by the worldline of the
splitting point $S$, which is characterized by the condition
$e_{(0)}\,^i=0$. A suitable construction of a tetrad frame, adapted to the
stationary character of the observer, is \cite{Maluf2}

\begin{equation}
e_{a\mu}=\pmatrix{
-A&0&0&-B\cr
0&1&0&0\cr
0&0&1&0\cr
0&0&0&C}\,,
\label{34}
\end{equation}
where

\begin{equation}
A=\biggl(1 -{H\over 2} \biggr)^{1/2}\,, \ \ \ \ 
AB={H\over 2}\,, \ \ \ \ \ 
AC=1 \,.
\label{35}
\end{equation}
In (34), $a$ and $\mu$ label rows and columns, respectively. It is not
difficult to verify that Eq. (34) yields \cite{Maluf2}

\begin{equation}
e_{(0)}\,^\mu=(1/A,0,0,0)\,,
\label{36}
\end{equation}
and

\begin{eqnarray}
e_{(1)}\,^\mu&=& (0,1,0,0)\,,  \nonumber \\
e_{(2)}\,^\mu&=& (0,0,1,0)\,,  \nonumber \\
e_{(3)}\,^\mu&=& (\,-H/(2A)\,,0,0,A). 
\label{37}
\end{eqnarray}
Note that if $H\ll 1$ we have $A\cong 1-H/4$ and therefore
$e_{(3)}\,^i=(0,0,A)\cong (0,0,1-H/4)$.

The frame is determined by fixing six conditions on $e_{a\mu}$. Equation (36)
fixes the kinematic state of the frame, since the three velocity conditions 
$e_{(0)}\,^i=0$ ensure that the frame is stationary. Three other conditions 
fix the spatial orientation of the frame. According to Eq. (37), 
$e_{(1)}\,^\mu$, $e_{(2)}\,^\mu$ and $e_{(3)}\,^\mu$ are unit vectors along 
the $x$, $y$ and $z$ axis, respectively. Note that by requiring $H=0$ we
obtain $e_a\,^\mu=\delta_a^\mu$, and consequently $T_{a\mu\nu}=0$.
The nonvanishing components of $T_{\mu\nu\lambda}$ are given in 
Ref. \cite{Maluf2}. They read

\begin{eqnarray}
T_{001}&=& {1\over 2} \partial_1 A^2\,, \ \ \ \ \ \ \ \
T_{002}= {1\over 2} \partial_2 A^2\,, \nonumber \\
T_{003}&=& {1\over 2} \partial_3 A^2 - A\partial_0 B\,, \ \ \ \ 
T_{013}= -A\partial_1 B\,, \nonumber \\
T_{023}&=& -A\partial_2 B\,, \ \ \ \  T_{301}= B\partial_1 A\,, \ \ \ \ \
T_{302}= B\partial_2 A\,, \nonumber \\
T_{303}&=& B\partial_3 A +{1\over 2}
\partial_0(-B^2+C^2)\,, \nonumber \\
T_{313}&=& {1\over 2} \partial_1(-B^2+C^2)\,, \ \ \ \ \ \
T_{323}= {1\over 2} \partial_2(-B^2+C^2)\,.
\label{38}
\end{eqnarray}
In the expressions above we have $(-B^2+C^2)=g_{33}$. 

The standard analysis of the 
breaking of parallelograms in a space-time with torsion is carried out by
considering two vectors, $A^\mu=dx^\mu$ and $B^\mu =\delta x^\mu$. The parallel
transport of $A^\mu$ along $\delta x^\mu$, and of $B^\mu$ along $dx^\mu$ are 
given by, respectively,

\begin{equation}
\delta A^\mu= -\Gamma^\mu_{\alpha\beta} A^\alpha \delta x^\beta\,,
\ \ \ \ \ 
\delta B^\mu= -\Gamma^\mu_{\alpha\beta} B^\alpha dx^\beta\,,
\label{39}
\end{equation}
where $\Gamma^\mu_{\alpha\beta}$ is an arbitrary  space-time connection, with
no {\it a priori} symmetry. The vectors 
$\lbrack A^\mu + (B^\mu +\delta B^\mu) \rbrack$
and $\lbrack B^\mu+ (A^\mu+\delta A^\mu)\rbrack$ do not form a closed 
parallelogram if the space-time is endowed with torsion (see Fig.2).
The breaking of a parallelogram was considered in Ref. 
\cite{Maluf1}, in the analysis of the Pound-Rebka experiment in the 
Schwarzschild space-time. In Fig. 2, the non-closure of the parallelogram is 
described by

\begin{eqnarray}
\lbrack A^\mu + (B^\mu +\delta B^\mu) \rbrack -
\lbrack B^\mu+ (A^\mu+\delta A^\mu)\rbrack 
&=&(\Gamma^\mu_{\alpha\beta}-
\Gamma^\mu_{\beta\alpha})dx^\alpha \delta x^\beta \nonumber \\
&=&T^\mu\,_{\alpha \beta}dx^\alpha \delta x^\beta\,.
\label{40}
\end{eqnarray}

\begin{figure}[h]
\centering
\includegraphics[width=0.60\textwidth]{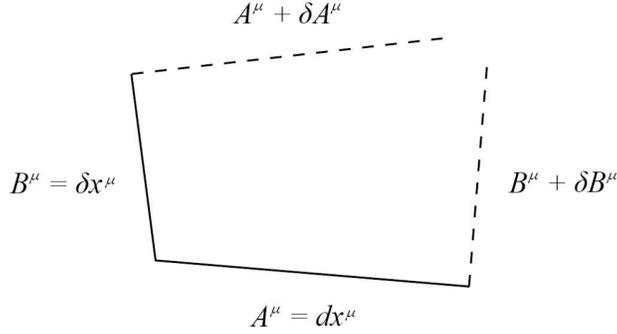}
\caption{The breaking of the parallelogram}
\end{figure}

We considered above the arm length $L$ to be
small compared to the width of the wave packet that describes the 
function $H$. 
This property was taken into account in Eqs. (31) and (32). By comparing Figs.
1 and 2 we are led to identify $a^\mu = dx^\mu$ and $v^\mu = \delta x^\mu$. 
Therefore, 

\begin{equation}
T^0\,_{\alpha\beta}\, dx^\alpha \delta x^\beta=
T^0\,_{\alpha\beta}\, a^\alpha v^\beta\,.
\label{41}
\end{equation}

A straightforward calculation leads to

\begin{equation}
T^0\,_{\alpha\beta}\, a^\alpha v^\beta=
T^0\,_{01}a^0v^1- T^0\,_{02} a^2 v^0-T^0\,_{12}a^2v^1\,.
\label{42}
\end{equation}
Taking into account the expressions of the torsion tensor components given by
Eq. (38), and using $g^{00}=(-1-{1\over 2}H)$ and 
$g^{03}=-{1\over 2}H$, we obtain

\begin{equation}
T^0\,_{\alpha\beta}\, a^\alpha v^\beta=
{L^2\over 4} \biggl[
{{ ( {{\partial H}\over {\partial x}})
-( {{\partial H}\over {\partial y}})}
\over
{  (1-{1\over 2}H)^{3/2}  }} 
\biggr]\,,
\label{43}
\end{equation}
and also $T^i\,_{\alpha\beta}\, a^\alpha v^\beta=0$, for $i=1,2,3$.
Equation (43) represents precisely the breaking of the parallelogram given
by Eq. (33). Note that Eq. (43) {\it cannot} be explained in terms of the 
curvature tensor, which depends on second order derivatives of the metric
tensor.

\section{The breaking of parallelograms in the $(+x,-x)$ and  
$(+x,-y)$ directions}

In this section we will consider two types of parallelograms,
geometrically distinct from the one depicted in Fig. 1, in the presence of
the gravitational wave described by Eqs. 
(19-21). The parallelogram of Fig. 1 is characterized as a $(+x,+y)$ 
parallelogram. In the geometrical constructions below, the vectors $v^\mu$
and $w^\mu$ are given as in Section 2, by Eqs. (24-25). Only the vectors 
$a^\mu$ and $b^\mu$ will be modified. The two new 
parallelograms are the following.

\subsection{Parallelogram in the $(+x,-x)$ directions}

The light beams travel back and forth in the positive $x$ direction 
($v^\mu$, $w^\mu$), and  in the negative $x$ direction ($a^\mu$, $b^\mu$), 
along arms of equal length $L$. Thus, the vectors $a^\mu$ and $b^\mu$ are 
defined by

\begin{equation}
a^\mu= \biggl( {L\over{(1-H/2)^{1/2}}},-L,0,0 \biggl)\,, \ \ \ \ \
H=H(x_0-L,y_0,t_0+dt)\,,
\label{44}
\end{equation}

\begin{equation}
b^\mu= \biggl( {L\over{(1-H/2)^{1/2}}},L,0,0 \biggl)\,, \ \ \ \ \
H=H(x_0,y_0,t_0+2dt)\,.
\label{45}
\end{equation}
The breaking of the parallelogram is given as before by 
$\Delta^\mu= (v^\mu+w^\mu)-(a^\mu+b^\mu) =  (\Delta^0,0,0,0)$.
After a simple algebra we find

\begin{equation}
\Delta^0={L^2\over 2}  \biggl[
{{  {{\partial H}\over {\partial x}} }
\over
{  (1-{1\over 2}H)^{3/2}  }} \biggr]
\,.
\label{46}
\end{equation}
The expression above is also obtained from the torsion tensor
components according to

\begin{equation}
T^0\,_{\alpha \beta}a^\alpha b^\beta=T^0\,_{01} a^0v^1+T^0\,_{10} a^1 v^0
=\Delta^0\,,
\label{47}
\end{equation}
where all quantities are given by eqs. (24), (38) and (44).

\subsection{Parallelogram in the $(+x,-y)$ directions}

Here we consider that the light beams travel in the positive $x$ direction 
($v^\mu$) and in the negative $y$ direction ($a^\mu$), along arms of
equal length $L$. The vectors $a^\mu$ and $b^\mu$ in this case are defined by

\begin{equation}
a^\mu= \biggl( {L\over{(1-H/2)^{1/2}}},0,-L,0 \biggl)\,, \ \ \ \ \
H=H(x_0,y_0-L,t_0+dt)\,,
\label{48}
\end{equation}

\begin{equation}
b^\mu= \biggl( {L\over{(1-H/2)^{1/2}}},0,L,0 \biggl)\,, \ \ \ \ \
H=H(x_0,y_0,t_0+2dt)\,.
\label{49}
\end{equation}
Once again the breaking of the parallelogram is given by 
$\Delta^\mu= (v^\mu+w^\mu)-(a^\mu+b^\mu) =  (\Delta^0,0,0,0)$. After some 
simple calculations we find 

\begin{equation}
\Delta^0=
{L^2\over 4} \biggl[
{{ ( {{\partial H}\over {\partial x}})
+( {{\partial H}\over {\partial y}})}
\over
{  (1-{1\over 2}H)^{3/2}  }} 
\biggr]\,.
\label{50}
\end{equation}
As in the previous case, the expression above is exactly obtained from
the torsion tensor components according to

\begin{equation}
T^0\,_{\alpha \beta}a^\alpha b^\beta=T^0\,_{01} a^0v^1+T^0\,_{20} a^2 v^0
=\Delta^0\,.
\label{51}
\end{equation}

Expressions (43), (47) and (51) are in agreement with the geometrical 
interpretation of the torsion tensor. The latter explains the geometrical 
breaking of parallelograms in space-time, and is fundamental in the 
description of the gravitational 
phenomena considered in the present framework. As we mentioned earlier, the 
breaking of the three parallelograms cannot be explained with the help of the 
curvature tensor of the space-time.

\section{Conclusions}

The breaking of the parallelogram depicted in Fig. 1 was obtained by 
considering first the physical vectors $v^\mu$, $w^\mu$, $a^\mu$ and $b^\mu$,
and then by analysing the parallel transport of vectors in a space-time 
endowed with torsion. The tetrad frame 
(34) and the torsion tensor components explain the breaking of the 
parallelogram. Therefore the measurement of the breaking of the 
parallelogram of Fig. 1 is, at the same time, a direct manifestation of 
gravitational waves and of the space-time torsion. This effect cannot be
explained directly by the metric tensor, and neither as a manifestation of the
curvature of the space-time. 

As a consequence of Eq. (46), we observe that two events that are simultaneous
in the flat space-time, may not be simultaneous in the presence of a 
plane-fronted gravitational wave. Two events are simultaneous in space-time if
the relevant parallelogram is not broken.

The quantity $\Delta^0$ given by Eq. (33) has dimension of length, and is 
measured on the worldline of the splitting point $S$. The interval of time 
elapsed between the arrival at the point $S$ of the two light rays, described 
by $w^\mu$ and $b^\mu$, is given by $\Delta \tau=\Delta^0/c$, 

\begin{equation}
\Delta \tau={L^2\over 4c} \biggl[
{{ ( {{\partial H}\over {\partial x}})
-( {{\partial H}\over {\partial y}})}
\over
{  (1-{1\over 2}H)^{3/2}  }}
\biggr]\,,
\label{52}
\end{equation}
where $c$ is the speed of light. 

We emphasize that we are considering a strictly thought experiment. However,
in the context of the LIGO experiment, we may obtain an expression for
$\Delta \tau$.
The length $L$ of one arm of the LIGO detector is $4\,km$. Assuming that the
light rays perform 200 round trips along each arm (after the splitting at the
point $S$ and before the detection), and also assuming the 
amplitude of the gravitational wave to be small, we find

\begin{equation}
\Delta \tau \approx (533\,m\cdot s) \biggl({{\partial H}\over {\partial x}}-
{{\partial H}\over {\partial y}} \biggr)\,.
\label{53}
\end{equation}
Note that since $H$ is a function of $x, y$ and 
$u=(z-t)/\sqrt{2}$, the partial derivatives are not linearly dependent on the
frequency of the incident gravitational wave. 

In view of (i) the lack of sufficient knowledge about the astrophysical 
source of gravitational perturbations, (ii) of the small intensity of the 
expected gravitational waves, and (iii) of the arbitrariness of the 
function $H(x,y,u)$, in practice it will be difficult to distinguish the 
detection of a gravitational wave signal from the background noise, at least
in the context of the LIGO experiment. The background
noise is characterized by a pattern of random oscillations that will be the
subject of precise measurements of the Fermilab Holometer experiment 
\cite{Hogan}. The latter is an experiment designed to study the properties of 
space-time at the smallest scales, by probing the nature of noise
in the very structure of the space-time. The Fermilab Holometer is constituted
by two similar interferometers, placed parallel and very close to each other,
so that the measurements in both interferometers may be correlated.
Although the function $H$ is not a priori known, we assume that it is given in
the form of a wave packet in the direction of propagation, and 
believe that its presence will momentarily alter the pattern of the 
background noise. Therefore, sudden and unexpected 
changes in the background noise may indicate the passage of a gravitational
wave. The Fermilab Holometer experiment might provide strong indications of 
the presence of non-linear waves.

\end{document}